\documentclass[twocolumn,twocolappendix]{openjournal}
\usepackage{graphics}
\usepackage{amssymb}
\usepackage{bbm}
\usepackage{amsmath,textcomp,array}
\usepackage{bm}

\lefthead{Heitmann et al.}
\righthead{The New Worlds Simulations}

\begin{document}


\title{The New Worlds Simulations: Large-scale Simulations across Three Cosmologies}

\author{Katrin Heitmann\altaffilmark{1}, Thomas Uram\altaffilmark{2}, Nicholas Frontiere\altaffilmark{3}, Salman Habib\altaffilmark{1,3}, Adrian Pope\altaffilmark{3}, Silvio Rizzi\altaffilmark{2}, Joe Insley\altaffilmark{3}}

\affil{$^1$HEP Division, Argonne National Laboratory, 9700 S. Cass Ave.,
 Lemont, IL 60439, USA}

\affil{$^2$Argonne Leadership Computing Facility, Argonne National Laboratory, Lemont, IL 60439, USA}

\affil{$^3$CPS Division, Argonne National Laboratory, 9700 S. Cass Ave.,
 Lemont, IL 60439, USA}

\begin{abstract}
In this paper we describe the set of ``New Worlds Simulations'', three very large cosmology simulations, Qo'noS, Vulcan, and Ferenginar, that were carried out on the Summit supercomputer with the Hardware/Hybrid Cosmology Code, HACC. The gravity-only simulations follow the evolution of structure in the Universe by each employing 12,288$^3$ particles in (3$h^{-1}$Gpc)$^3$ volumes, leading to a mass resolution of $m_p\sim 10^9h^{-1}$M$_\odot$. The simulations cover three different cosmologies, one $\Lambda$CDM model, consistent with measurements from Planck, one simulation with massive neutrinos, and one simulation with a varying dark energy equation of state. All simulations have the same phases to allow a detailed comparison of the results and the investigation of the impact of different cosmological parameters. We present measurements of some basic statistics, such as matter power spectra, correlation function, halo mass function and concentration-mass relation and investigate the differences due to the varying cosmologies. Given the large volume and high resolution, these simulations provide excellent bases for creating synthetic skies.  A subset of the data is made publicly available as part of this paper.
\end{abstract}


\section{Introduction}

Future cosmological surveys, such as the Legacy Survey of Space and Time (LSST) to be carried out with the Rubin Observatory~\citep{lsst}, and newly developed instruments and missions, such as the Dark Energy Spectroscopic Instrument (DESI, \citealt{desi}), Euclid~\citep{euclid}, and Roman~\citep{roman},  aim to shed light on the nature of the dark Universe and unravel the mysteries of the origin of dark matter and the accelerated expansion of the Universe.  Measurements from these surveys have reached a level of accuracy that they can also provide constraints on neutrino masses. To interpret the wealth of data, simulations have become ever more important. They provide predictions in the nonlinear regime of structure formation which holds key information about cosmology. They also function as testbeds for complex analysis and processing tasks as well as for the exploration of possible new cosmological probes. For many of these tasks, large simulation volumes ($\sim$ (3$h^{-1}$Gpc)$^3$ or larger) with excellent mass resolution (particle masses of $m_p\sim 10^9h^{-1}$M$_\odot$ or better) are required to capture survey size and depth. 

\begin{figure}[t]
\includegraphics[width=3.5in]{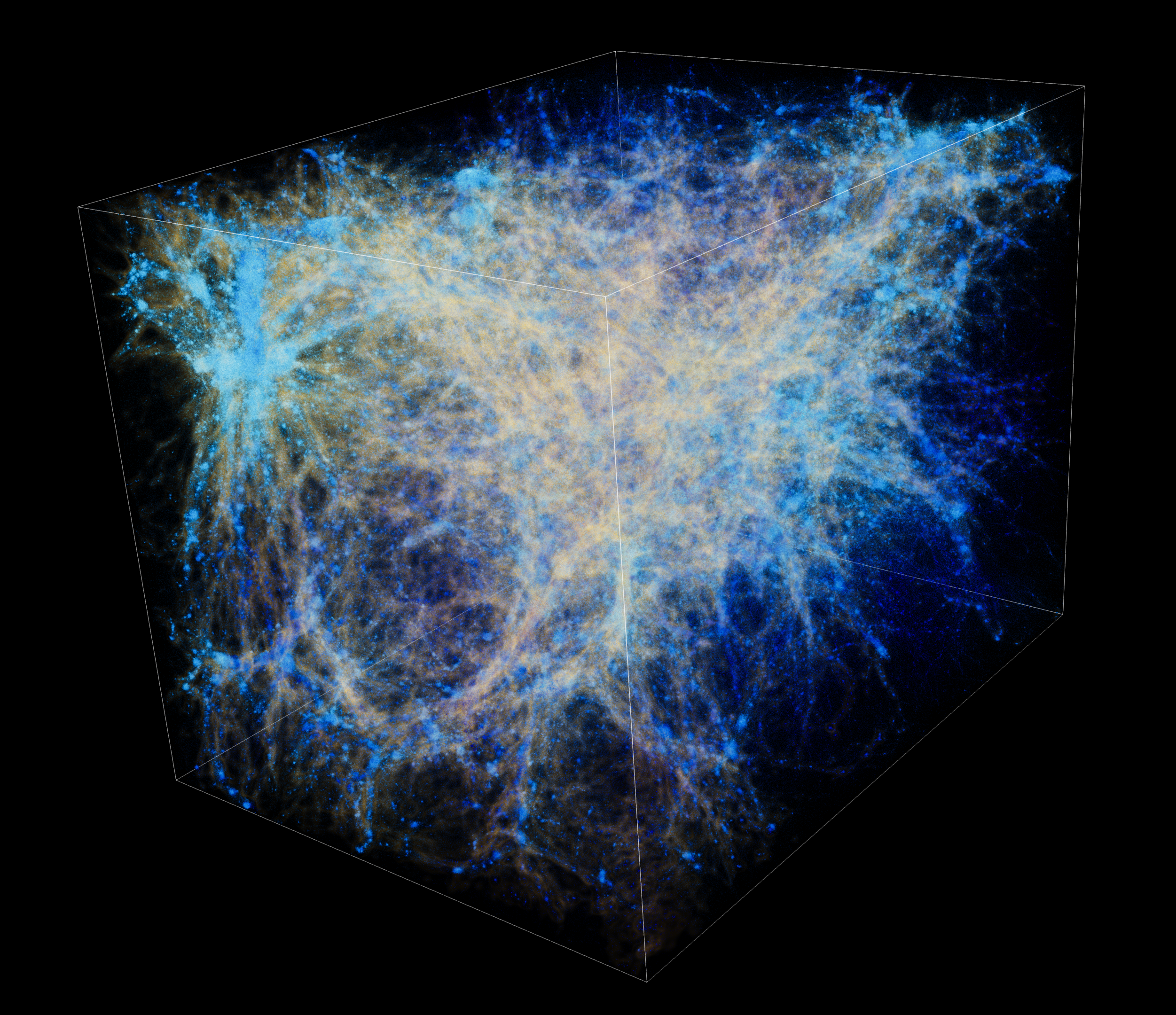}
\caption{Visualization of the particle density from Rank 0 (out of 24576 ranks) from the Ferenginar simulation at $z=0$.} 
\label{fig:qonos_vis}
\end{figure}

Large-scale simulations with high mass resolution have become increasingly important for cosmological surveys to prepare for data arrival. For example, the LSST Dark Energy Science Collaboration (LSST DESC) \citep{desc} has created the second data challenge (DC2)~\citep{dc2} based on the Outer Rim simulation \citep{heitmann19a}. DC2 covers 400 deg$^2$ of the LSST footprint and five years of observations. The Outer Rim simulation is based on a best-fit WMAP-7 cosmology and has been now updated by the Last Journey simulation, featuring a cosmological model close to the best-fit Planck results \citep{planck18}. Another example includes the Euclid Flagship simulation~\citep{potter} to model the ongoing Euclid survey. In addition, major simulation suites have been created to support survey science, including the Buzzard and MICE simulations for the Dark Energy Survey (DES)~\citep{2019arXiv190102401D,2015MNRAS.453.1513C}, and the {\sc AbacusSummit} suite~\citep{abacus} for DESI science. 

In this paper, we introduce three new simulations, similar in size to the Outer Rim, Last Journey and Euclid Flagship simulations. The Qo'noS, Vulcan, and Ferenginar simulations were carried out with the Hardware/Hybrid Accelerated Cosmology Code (HACC), described in~\cite{habib14}.
Each simulation evolves 12,288$^3$ particles in a (3$h^{-1}$Gpc)$^3$ volume. The models feature three different cosmologies, a $\Lambda$CDM model with the current best-fit Planck values (Qo'noS), the same cosmology with a massive neutrino component added (Vulcan), and a dynamical dark energy model, with a dark energy equation of state parameterized by $(w_0,w_a)$ (Ferenginar).  Figure~\ref{fig:qonos_vis} shows the visualization of the data from one rank from the Ferenginar simulation at $z=0$. All simulations have the same initial phases, enabling detailed comparisons of subtle differences that are not washed out due to uncertainties from cosmic variance. The set of simulations also lends itself well to a blind analysis challenge to investigate how sensitive cosmological probes are to small changes in cosmology. 

The creation of such large simulations poses a challenge with regard to code scalability, as both the computational and memory requirements can only be satisfied by scaling to large supercomputers. We have developed HACC for more than a decade and have optimized the code to run on the fastest supercomputers in the world available at any given time. The architectures have changed considerably during that time span, including hardware-accelerated machines like Roadrunner\footnote{https://en.wikipedia.org/wiki/Roadrunner\_(supercomputer)}, Titan\footnote{https://en.wikipedia.org/wiki/Titan\_(supercomputer)} and Summit\footnote{https://en.wikipedia.org/wiki/Summit\_(supercomputer)} as well as many-core architectures like Mira\footnote{https://www.alcf.anl.gov/alcf-resources/mira}, Sequoia\footnote{https://en.wikipedia.org/wiki/Sequoia\_(supercomputer)} and Theta\footnote{https://www.alcf.anl.gov/alcf-resources/theta}. 
HACC has run successfully on all of these machines and has been recently prepared to run on the exascale platforms Frontier and Aurora at the Leadership Computing Facilities at Oak Ridge and Argonne, respectively. For each simulation described in this paper, we utilized 4096 nodes on Summit, close to the full machine. A set of first level analysis tasks was carried out on the fly. These tasks include halo finding with different halo definitions, power spectrum estimates, and halo core finding to identify substructures. Results from analyzing these first level data products are presented in this paper and the differences due to the cosmological parameter choices are highlighted.

The paper is organized as follows. In Section~\ref{setup} we provide an overview of the three simulations and the outputs that were stored. Next, in Section~\ref{results} we present measurements for a range of statistics and show the deviations due to the differences in the cosmological models simulated. We briefly discuss publicly available data products in Section~\ref{sec:release}, and conclude in Section~\ref{summary}.

\section{Simulation Set-up and Data Products}
\label{setup}
In this section we provide details about the set-up of the simulations and data products generated from the simulation runs. 

\subsection{Cosmological Models}

The three simulations introduced in this paper, Qo'noS, Vulcan, and Ferenginar each evolve 12,288$^3$ particles in a (3$h^{-1}$Gpc)$^3$ volume. We choose the same total matter density value, baryon density value, Hubble parameter, $\sigma_8$ and spectral index for all three cosmologies as follows:
\begin{eqnarray}
\label{eq:param1}
\Omega_{m}&=&0.3096,\\
\omega_b&=&0.02242,\\ 
h&=&0.6766,\\ 
\sigma_8&=&0.8102,\\
n_s&=&0.9665. \label{eq:param5}
\end{eqnarray}
In addition, we restrict the simulations to flat universes with $\Omega_k=0$ and hence $\Omega_\Lambda = 1-\Omega_m$. These parameters lead to a particle mass of $m_p=1.26\cdot10^9h^{-1}$M$_\odot$, the same for all simulations. The force resolution employed is 2.4$h^{-1}$kpc. The simulations were started at $z=200$ using the Zel'dovich approximation~\citep{Zel70} to assign the initial particle positions and velocities. The input transfer functions were generated using CAMB~\citep{camb}. Each simulation has the same initial phases, and were individually run on 4096 nodes on Summit using forty two cores and all six GPUs per node. This enables detailed comparisons of the results and investigations of the differences due to changes of cosmological parameters. 

in addition to the five parameters given in Eqs.~(\ref{eq:param1}) -- (\ref{eq:param5}), for the Qo'noS simulation we have chosen a constant dark energy of state and massless neutrinos:
\begin{eqnarray}
w_0&=&-1,\\
w_a&=&0,\\
\Omega_\nu&=&0.
\end{eqnarray}

The Vulcan simulation is similar to Qo'noS with the addition of a small massive neutrino component, but with $\Omega_m$ unchanged, so that $\Omega_{cdm}=0.25833$ (a slightly lower value than for Qo'noS). 
We treat massive neutrinos at the background level
instead of simulating them as a separate particle species. This approach avoids numerical inaccuracy induced by simulating particle populations with very different masses. A detailed description of our approach is given in Upadhye et al. \cite{2014PhRvD..89j3515U}
and \cite{2016ApJ...820..108H}. Its validity with regard to
power spectrum measurements on large to quasi-linear
scales is discussed extensively in \cite{2014PhRvD..89j3515U}
and the effect of neutrinos on the mass function is investigated for two models in \cite{2019arXiv190110690B}, including one for high-mass neutrinos. We chose a model model the neutrino mass of three species via their sum to be $\sum m_\nu=0.1$eV. This leads to the following parameter choices (where we have used the relationship $\Omega_\nu=\sum m_\nu/93.1{\rm eV}h^2$): 
\begin{eqnarray}
w_0&=&-1,\\
w_a&=&0,\\
\Omega_\nu&=&0.00234.
\end{eqnarray}

Finally, the Ferenginar simulation employs the same cosmology as the Qo'noS simulation but where we replace the cosmological constant with a dynamical dark energy model, parameterized via 
\begin{equation}
\label{eq:wz}
w(a)=w_0+w_a(1-a).
\end{equation}
The three additional parameters to Eqs.~(\ref{eq:param1}) -- (\ref{eq:param5}) are given by:
\begin{eqnarray}
w_0&=&-0.9,\\
w_a&=&0.7,\\
\Omega_\nu&=&0.
\end{eqnarray}

To gain an intuitive understanding about the three different models and how we expect structure-formation results to differ, we briefly discuss the resulting Hubble expansion rates $H(z)$ for the Qo'noS and Vulcan simulations compared to the Ferenginar simulation. The Hubble expansion rate for a dark energy model with an equation of state parameterized via Eq.~(\ref{eq:wz}) is given by
\begin{equation}
\begin{split}
H(z)&=H_0\biggl\{\Omega_m(1+z)^3+\Omega_\Lambda(1+z)^{3(1+w_0+w_a)}
    \\
 &\qquad\quad \times\exp\left\{-\frac{3w_a z}{1+z}\right\}
  \biggr\}^{\!1/2},
\end{split}
\end{equation}
which simplifies for $\Lambda$CDM cosmologies (Qo'noS and Vulcan) to
\begin{equation}
H(z)=H_0\biggl\{\Omega_m(1+z)^3+\Omega_\Lambda
  \biggr\}^{\!1/2}.
\end{equation}

\begin{figure}[t]
\includegraphics[width=3.5in]{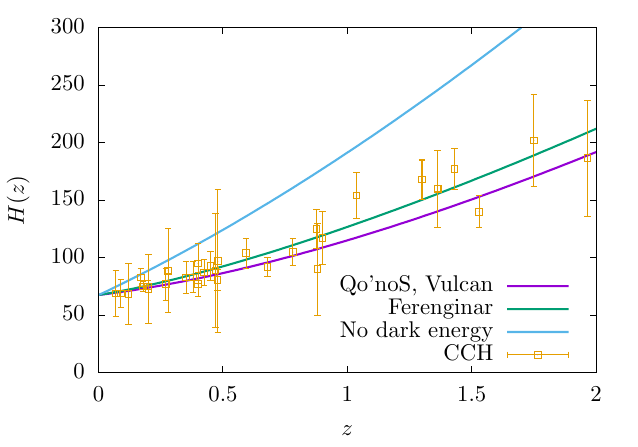}
\caption{Redshift evolution of the Hubble parameter for the different cosmologies considered in the paper and a model without dark energy for comparison (purple: $\Lambda$CDM, green: dynamical dark energy, blue: no dark energy). Additionally, we show data from cosmic chronometers (CCH), provided in~\cite{CCH}, Table~1. } 
\label{fig:hz}
\end{figure}

Figure~\ref{fig:hz} shows the evolution of the Hubble parameter $H(z)$ as a function of redshift for the three models considered in this paper. For comparison, we also show a model with no dark energy, i.e. $\Omega_m=1$ and 31 measurement points based on cosmic chronometers (CCH), provided in Table~1 in~\cite{CCH}. We chose this data set due to its large redshift coverage, out to $z\sim 2$. The measurements illustrate the uncertainties at higher redshifts that still hold clues about the exact nature of the cause for the accelerated expansion of the Universe. The evolution of the Hubble parameter in our dynamical dark energy model is steeper at early times than for a $\Lambda$CDM model but more comparable at lower redshifts due to choice of the same value for $H(0)$ for the models considered.  

\begin{figure}[t]
\includegraphics[width=3.5in]{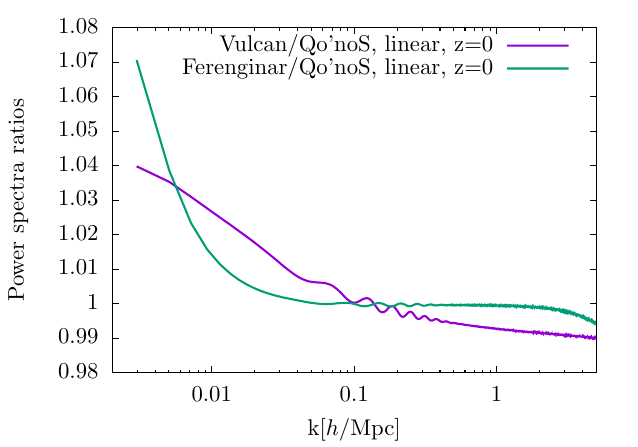}
\caption{Ratios of the power spectra from the initial conditions of the three simulations, all scaled to $z=0$ via their linear growth functions. We use the $\Lambda$CDM model as base. For Qo'noS and Ferenginar, the exact, scale-independent growth function was used and the full matter power spectrum is shown. For Vulcan, only the $bc$ component of the power spectrum is shown and the scale dependence induced by the neutrinos was not taken into account when scaling to $z=0$.} 
\label{fig:pk_ratio_z0}
\end{figure}

Another measurement that provides insights into the differences of the models considered is the linear power spectrum. Figure~\ref{fig:pk_ratio_z0} shows the ratio of the initial power spectra from the three simulations, scaled to $z=0$ using the growth functions corresponding to each model. Since for all three models we chose the same value for $\sigma_8$, the power spectra cross at the scale corresponding to the normalization. Both the dynamical dark energy model and the neutrino model have more power on large scales compared to the $\Lambda$CDM cosmology. The differences on small scales are minimal. The figure demonstrates the major challenge that we will face when trying to disentangle subtle physics effects such as small neutrino masses or deviations from $w=-1$ from next generation data. Simulations as those presented here will be important to demonstrate that our cosmological probes are indeed sensitive enough for this endeavour. 

\begin{figure*}[t]
\centering
\includegraphics[width=6.5in]{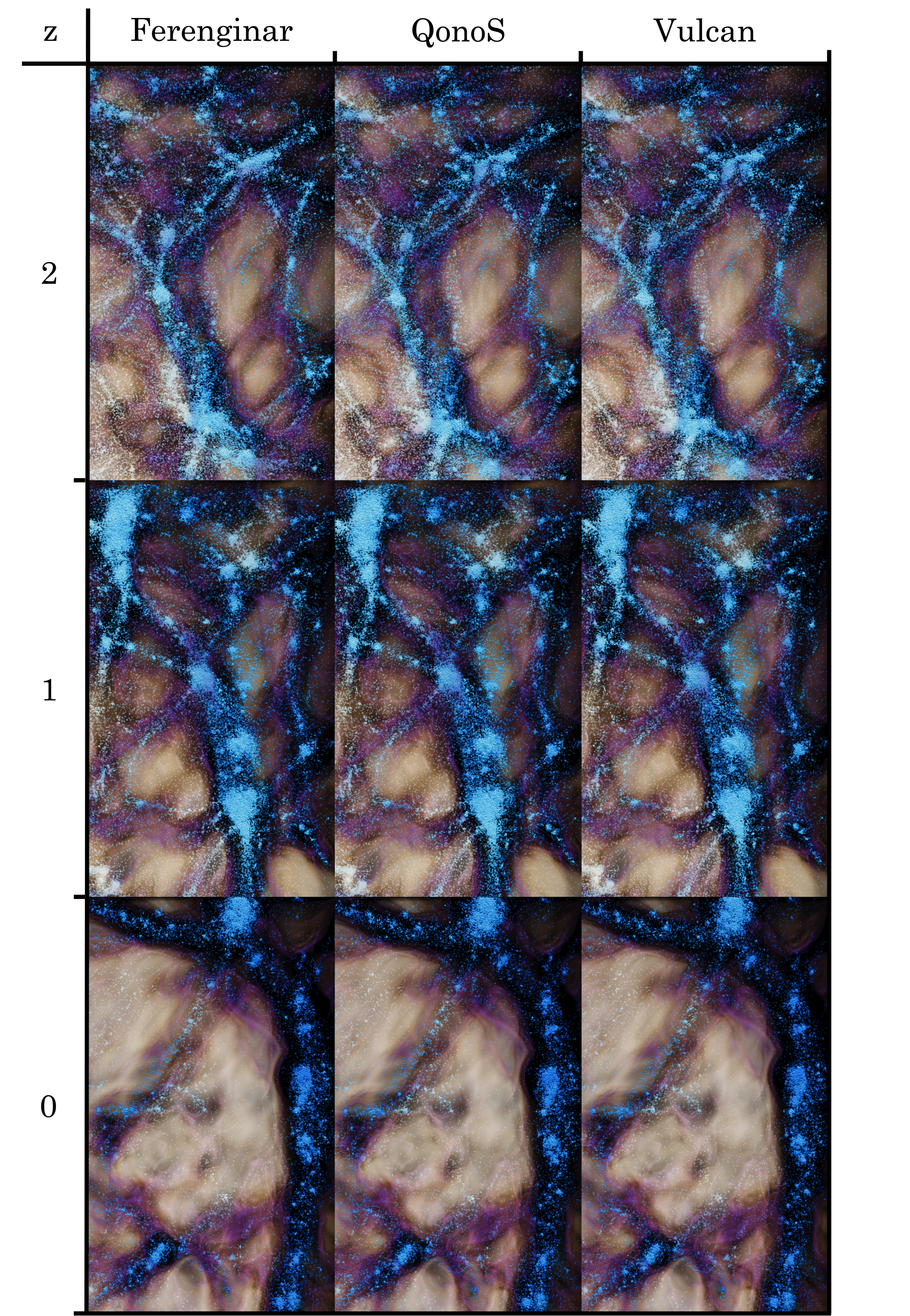}
\caption{Zoom-in visualization at three redshifts $z=2,1,0$ (top to bottom) for the three different simulations. Particles are overlayed on the density representation. Differences are clearly visible at early times, the Ferenginar simulation shows tighter and larger structures in the high-density regions. Over time, the structures appear more and more similar.} 
\label{fig:zoom-in_vis}
\end{figure*}

Finally, we provide some visual insights into the different cosmologies in Fig.~\ref{fig:zoom-in_vis}. We show a zoom-in for the outputs of one of the ranks from the simulations. The camera position in these ParaView\footnote{https://www.paraview.org/} images is the same for all three models and the different redshifts shown. The center column shows the result for the $\Lambda$CDM cosmology (Qo'noS), on the left, we show the varying dark energy equation of state simulation (Ferenginar) and on the right the cosmology that includes neutrinos (Vulcan). From top to bottom we show three redshifts, $z=2,1$ and $z=0$. At earlier times (first row), differences between Ferenginar and Qo'noS/Vulcan (center/right) are more apparent. For example, the large nodal structure in the upper right corner is tighter in the Ferenginar simulation as is the structure in the lower half of the figure close to the right edge of the figure. Moving forward in time (second row at $z=1$ and third row at $z=0$) the differences are less pronounced. The visualizations also show how the void structures grow over time and the filaments become more defined as expected. Throughout Section~\ref{results} we will quantitatively show how the three simulations display different clustering properties early on and how the differences reduce over time.

\subsection{Outputs}
The data sizes for the simulations are very large. For example, a single time snapshot, containing positions and velocities, a particle ID and a potential value for each particle, has a size of close to 70TB (after lossless compression). This makes it prohibitive to save the full raw particle data output across all the time steps. To reduce the data size to a manageable amount, we ran all first level analysis on the fly (halo finding, power spectrum measurements, downsampling of particle snapshots) and did not save full raw data snapshots. 
We have analyzed 101 time snapshots between $z=10$ and $z=0$, evenly spaced in $\log_{10}(a)$ and saved a diverse set of analysis products at the following output values in $z$:
\begin{eqnarray}
z&=&\left\{10.04, 9.81, 9.56, 9.36, 9.15, 8.76, 8.57, 8.39, 8.05,  \right.\nonumber\\
&&7.89, 7.74, 7.45, 7.31, 7.04, 6.91, 6.67, 6.56, 6.34,   \nonumber\\
&&6.13, 6.03, 5.84, 5.66, 5.48, 5.32, 5.24, 5.09, 4.95,   \nonumber\\
&&4.74, 4.61, 4.49, 4.37, 4.26, 4.10, 4.00, 3.86, 3.76,  \nonumber\\
&&3.63.  3.55, 3.43, 3.31, 3.21, 3.10, 3.04, 2.94, 2.85,   \nonumber\\
&&2.74, 2.65, 2.58, 2.48, 2.41, 2.32, 2.25, 2.17, 2.09,   \nonumber\\
&&2.02, 1.95, 1.88, 1.80, 1.74, 1.68, 1.61, 1.54, 1.49,   \nonumber\\
&&1.43, 1.38, 1.32, 1.26, 1.21, 1.15, 1.11, 1.06, 1.01,   \nonumber\\
&&0.96, 0.91, 0.86, 0.82, 0.78, 0.74, 0.69, 0.66, 0.62,   \nonumber\\
&&0.58, 0.54, 0.50, 0.47, 0.43, 0.40, 0.36, 0.33, 0.30,  \nonumber\\
&& 0.27, 0.24, 0.21, 0.18, 0.15, 0.13, 0.10, 0.07, \nonumber\\
&&\left.  0.05, 0.02, 0.00\right\}.
\end{eqnarray}
For these redshifts we save halo catalogs based on the Friends of Friends (FOF) definition~\citep{davis85} using a linking length $b=0.168$. For FOF halos with more than 500 particles, we measure the mass overdensity in spheres around the potential minimum center until it falls below 200 times $\rho_c=3H^2/8\pi G$, defining the radius of the SO halos. Then we determine the enclosed overdensity mass M$_{200c}$.

We also saved all FOF halo particles. We store FOF halos down to 20 particle halo size, while SO halos have at least 500 particles per halo. In addition to a range of halo properties, we also saved SO profiles. To enable substructure tracking, we stored halo core particle data (for more details about the core concept, see \citealt{2020arXiv200808519R} and \citealt{2023OJAp....6E..24K} for an implementation example concerning galaxies in clusters). Finally, we saved down-sampled particle at 1\% snapshots for 59 steps, starting at  $z=3.04$, following the list above.

\section{Results in the Nonlinear Regime}
\label{results}

Next, we display measurements comparing the simulations in the nonlinear regime, showing results for both particle- and halo-based statistics. We compare some of the predictions with recently released emulators to verify the accuracy of the simulations.  

\subsection{Power Spectra}

\begin{figure}[t]
\centerline{\includegraphics[width=3.5in]{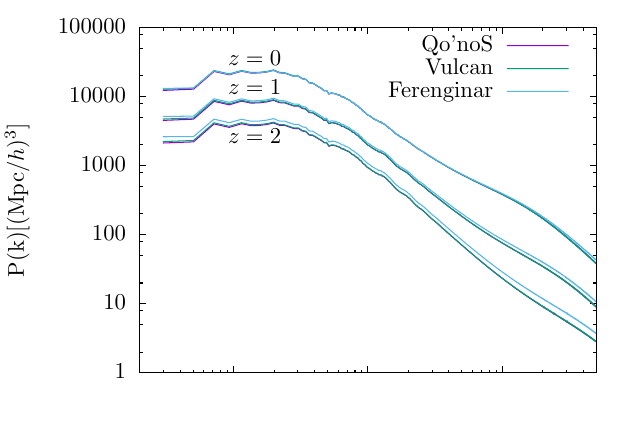}}
\vspace{-1.35cm}
\centerline{\hspace{0.45cm}\includegraphics[width=3.3in]{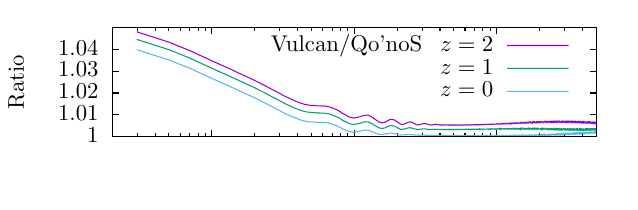}}
\vspace{-1.3cm}
\centerline{\hspace{0.45cm}\includegraphics[width=3.3in]{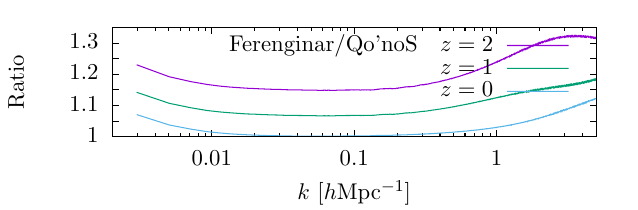}}
\vspace{0.1cm}
\caption{Upper panel: Power spectra at three redshifts, $z=0,1$ and $z=2$. For Qo'noS and Ferenginar, the full matter power spectrum is shown while for Vulcan, we show only the $bc$ component of the power spectrum. The difference between Qo'noS and Vulcan is very small. The difference with respect to Ferenginar decreases at lower redshift. Middle and lower panel: Ratio of power spectra for the same three redshifts. The $\Lambda$CDM simulation (Qo'noS) is used as base in all cases. } 
\label{fig:pk_nl_z0}
\end{figure}

We begin with measurements of the matter power spectra shown in the upper panel of Fig.~\ref{fig:pk_nl_z0} at three redshifts, $z=0,1,2$. Similar to the linear power spectra shown in Fig.~\ref{fig:pk_ratio_z0}, we plot the $bc$ component only for the Vulcan run and the full matter power spectra for the Qo'noS and Ferenginar simulations. The nonlinear evolution for the $\Lambda$CDM (Qo'noS) and the neutrino simulation (Vulcan) are very similar, while the dynamical dark energy model in the Ferenginar run exhibits larger nonlinear power at small scales compared to the other two simulations.
All three cosmological models are normalized with respect to the same value for $\sigma_8$ which leads to closer agreement at $z=0$ compared to the higher redshifts. The middle panel of Fig.~\ref{fig:pk_nl_z0} shows the ratio of the Vulcan and Qo'noS simulation for the three redshifts and the lower panel shows the ratio of the Ferenginar and Qo'noS simulations. The difference between the $bc$ component of the neutrino simulation and the $\Lambda$CDM simulation is larger on large scales (small $k$ values) while for the dynamical dark energy, there is also a difference at small scales. The dynamical dark energy model exhibits more power than the other two models. The excess power for the Ferenginar run at early redshifts manifests itself in other structure formation measurements as we will show later.

\begin{figure}[t]
\centerline{\includegraphics[width=3.5in]{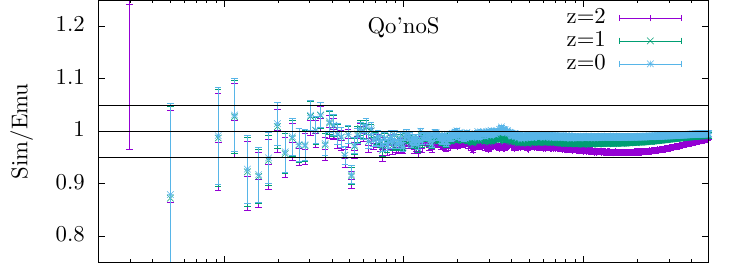}}
\centerline{\includegraphics[width=3.5in]{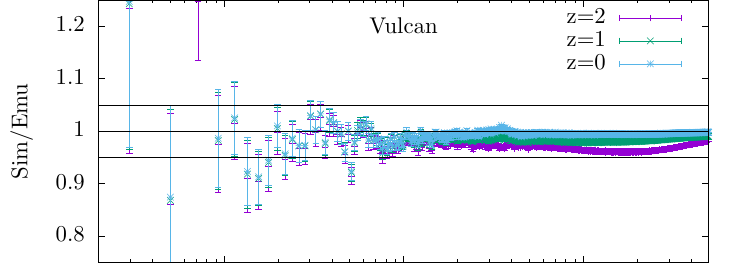}}
\centerline{\includegraphics[width=3.5in]{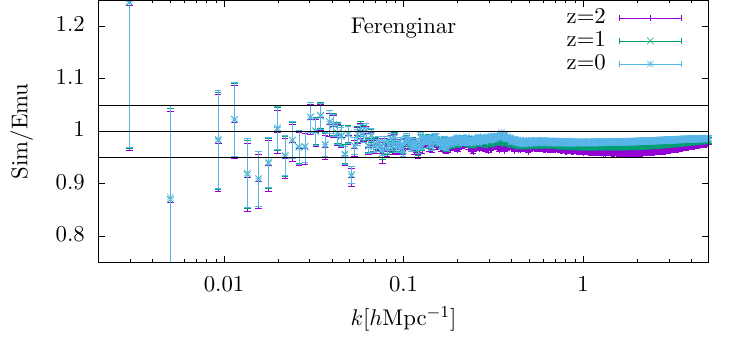}}
\vspace{-0.15cm}
\caption{Comparison of the simulation results to the CosmicEmu introduced in \cite{Moran:2022iwe}. The upper panel shows the comparison of the emulator to the Qo'nos simulation, the middle panel the Vulcan results and the last panel shows the Ferenginar results. In all three cases, three redshifts are shown (purple: $z=2$, light blue: $z=1$ and light green: $z=0$). The best agreement is achieved for $z=0$ at better than 1\% throughout the $k$-range considered. The agreement degrades with higher redshifts (4\% at $z=2$) which is consistent with the findings in \cite{Moran:2022iwe}. The error bars represent the uncertainties due to cosmic variance.} 
\label{fig:pk_emu}
\end{figure}

In Fig.~\ref{fig:pk_emu} we compare the simulation results at three redshifts with the CosmicEmu, an  emulator released in~\cite{Moran:2022iwe}\footnote{https://github.com/lanl/CosmicEmu} and based on the Mira-Titan Universe simulations~\citep{2016ApJ...820..108H}. Using emulators to predict cosmological measurements from simulations over a wide range of cosmologies has become a commonly used approach after it was introduced in the cosmological context in \cite{2006ApJ...646L...1H,2007PhRvD..76h3503H}. The emulator schemes include a wide range of predictions, for e.g. matter power spectra \citep{2010ApJ...713.1322L,2014ApJ...780..111H,2019MNRAS.484.5509E,2021MNRAS.507.5869A,Moran:2022iwe}, mass functions \citep{2019ApJ...872...53M,emu_massf}, and galaxy correlation functions \citep{2019ApJ...874...95Z,2023ApJ...952...80K}. The major advantage of emulators over fitting functions such as Halofit \citep{2003MNRAS.341.1311S} for the power spectrum or the Tinker mass function prediction \citep{2008ApJ...688..709T} is their high accuracy, even away from $\Lambda$CDM. In our comparison here, the accuracy of the power spectrum prediction is at the $1\%$ level at $z=0$ for all three simulations, covering $\Lambda$CDM, neutrinos and a varying dark energy equation of state. Overall, the simulations and emulator predictions across models and redshifts are in agreement at better than 5\% out to $k\sim 5 h$Mpc$^{-1}$, consistent with the results reported in~\cite{Moran:2022iwe}. The error bars shown in Fig.~\ref{fig:pk_emu} are derived by assuming a Gaussian Random Field with the measured power spectrum where the standard deviation of power in the $i$th wavenumber bin is given by $\sigma_{P(k_i)} = {P(k_i)}/{\sqrt{N_{\rm modes}(k_i)}}$.

\subsection{Matter Correlation Function}

\begin{figure}[t]
\includegraphics[width=3.5in]{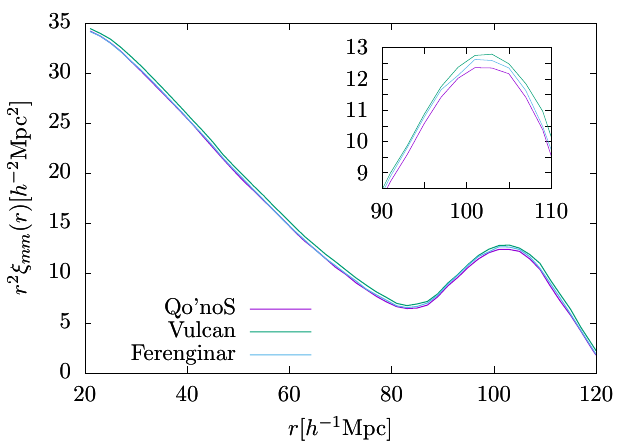}
\caption{Matter correlation functions for all three models at $z=0$. The small inset image shows the difference for the three models at the BAO peak location. Overall, at this redshift, the differences between the three models are minimal. } 
\label{fig:xi_mm}
\end{figure}

Following on, we show results for the two-point matter correlation function, $\xi_{mm}(r)$, at $z=0$ in Fig.~\ref{fig:xi_mm}, the equivalent of the power spectrum measurement in real space. We follow the same strategy to measure $\xi_{mm}(r)$ as described in \cite{2021ApJS..252...19H} and refer the reader to that publication for details. To make the evaluation of the correlation function from the particles computationally feasible, we only use 0.01\% of the full particle data set. We focus on the region of the baryon acoustic oscillation peak as this is the major target of ongoing spectroscopic surveys such as DESI (Fig.~\ref{fig:xi_mm} shows in the inset the BAO peak region).  As for the power spectra, at $z=0$ the results for the different models are very similar and the measurement for the Vulcan simulation only includes the CDM and baryon components of the simulation. On all scales, the Qo'noS simulation has the lowest amplitude, followed by the Ferenginar simulation. The Vulcan results are the highest. On smaller scales before the BAO peak ($20h^{-1}$Mpc$< r <$ 80$h^{-1}$Mpc) the Qon'oS and Ferenginar results are very similar, while the amplitude of the Vulcan result is higher at the 1-2\% level.

\subsection{Halo-Matter Correlation Function}

\begin{figure}
\centerline{\includegraphics[width=3.5in]{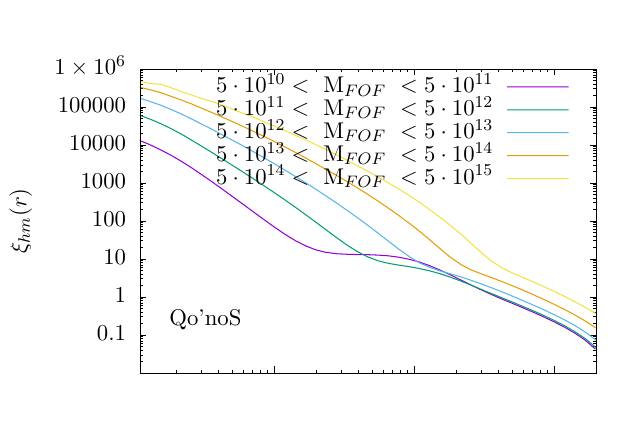}}
\vspace{-1.95cm}
\centerline{\includegraphics[width=3.5in]{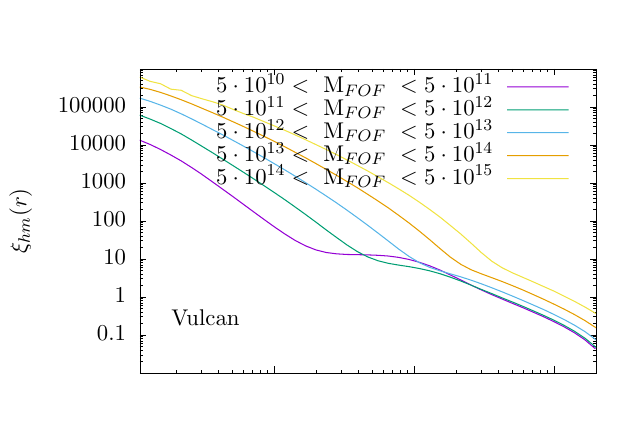}}
\vspace{-1.95cm}
\centerline{\includegraphics[width=3.5in]{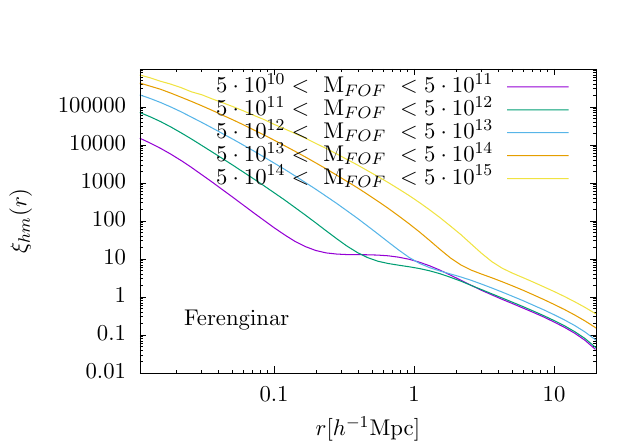}}
\caption{Halo-matter correlation function for five mass bins, separately shown for each model at $z=0$. } 
\label{fig:xi_hm}
\end{figure}

\begin{figure}
\centerline{\includegraphics[width=3.5in]{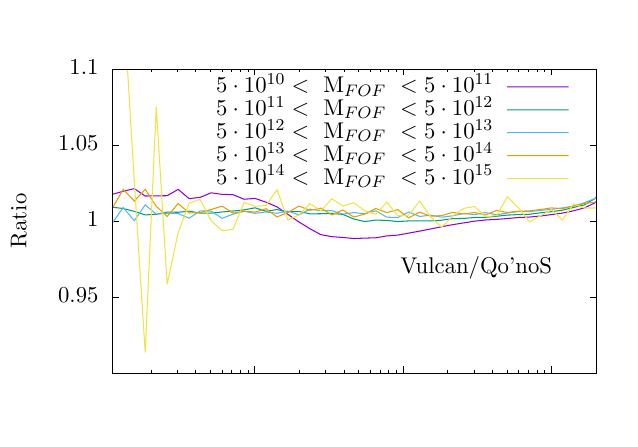}}
\vspace{-1.95cm}
\centerline{\hspace{0.2cm}\includegraphics[width=3.4in]{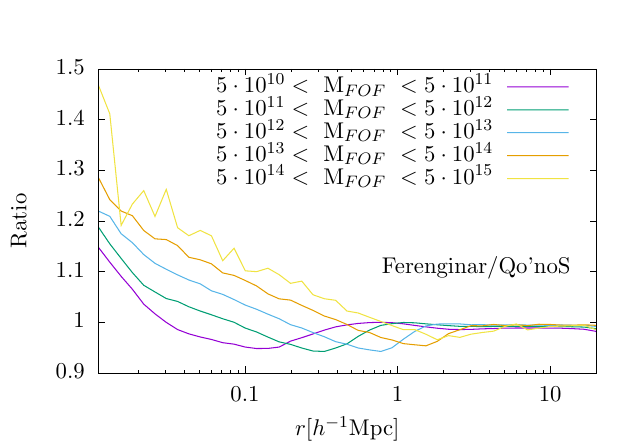}}
\caption{Ratios of the halo-matter correlation function for five mass bins at $z=0$. Upper plot: Ratio between Vulcan and Qo'noS, lower plot: ratio between Ferenginar and Qo'nos. The effect of the neutrinos in the Vulcan simulation is minimal while the effect of the dynamical dark energy in Ferenginar changes the halo-matter correlation by up to 30\% in the largest halo mass bin at small distances. } 
\label{fig:xi_hm_ratio}
\end{figure}

The halo-matter correlation functions for all three models are shown in Fig.~\ref{fig:xi_hm} for five different halo mass bins at redshift $z=0$. We use the FOF halo sample for this measurement. The ratios for Vulcan and Ferenginar with respect to the Qo'noS result are shown in Fig.~\ref{fig:xi_hm_ratio}. The difference between the two $\Lambda$CDM simulations with (Vulcan) and without neutrinos (Qo'noS) shown in the upper panel, is very small, at the percent level. In the largest halo mass bin there is some noise on small scales due to the limited number of halos. The differences between the $\Lambda$CDM and the dynamical dark energy model (Ferenginar) are small on large scales (down to $\sim 1 h^{-1}$Mpc) and increase on small scales up to 30\%. Again, the mass bin with the cluster sized objects is noisy due to the small number of halos. On larger scales, beyond $r=2h^{-1}$Mpc the differences between the halo-matter correlation functions are minimal for the different cosmologies.    

\subsection{Halo Mass Function}

\begin{figure}[t]
\centerline{\includegraphics[width=3.5in]{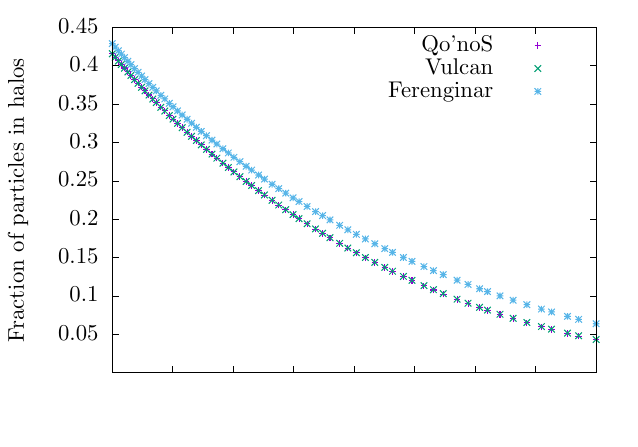}}
\vspace{-1.5cm}
\centerline{\includegraphics[width=3.5in]{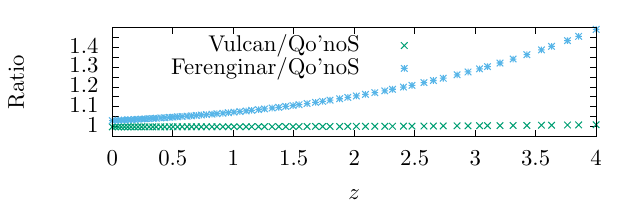}}
\caption{Upper panel: Evolution of the number of particles in halos. Shown is the fraction of particles in halos as a function of redshift out to $z=4$. The Qo'nos and Vulcan simulation have very similar fractions, while the Ferenginar simulation has a larger fraction of particles reside in halos, in particular at earlier times. Lower panel: Ratio of the number of particles in halos with respect to the $\Lambda$CDM model, Qo'noS.} 
\label{fig:haloc}
\end{figure}

Before we show results for the halo mass function, we compare the number of particles in halos as a function of redshift $z$ in Fig.~\ref{fig:haloc}. The halo model assumes that all particles reside in halos. In a simulation, due to the finite mass resolution, this is not the case. It is interesting, however, to investigate what fraction of particles do reside in halos and how this fraction depends on cosmology. This measurement is very similar to the halo mass function in spirit but has in addition information about the time evolution. However, this measurement does not contain information about the number of halos in different mass bins. The upper panel in Fig.~\ref{fig:haloc} shows the direct comparison of the three simulations. The difference between Qo'noS ($\Lambda$CDM) and Vulcan ($\Lambda$CDM + massive neutrinos) is very small, the points are basically on top of each other. For the Vulcan simulation, we do not include neutrino particles since our simulation approach only evolves the $bc$ component of the matter in the simulation. The dynamical dark energy model (Ferenginar, light blue) exhibits more particles in halos at earlier redshift -- this is consistent with the excess in the nonlinear power spectrum shown in Fig.~\ref{fig:pk_nl_z0}. At later stages the difference decreases. The lower panel measures the ratio of the simulations, showing that indeed the Ferenginar simulation has early on (at $z=4$) 40\% more particles in halos than the $\Lambda$CDM models, with the ratio declining steadily until reaching $z=0$. 

\begin{figure}[t]
\centerline{\includegraphics[width=3.5in]{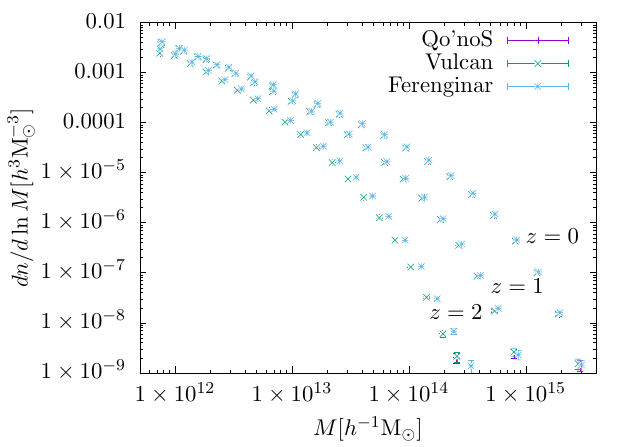}}
\caption{Halo mass function for $M_{200c}$ at three different redshifts for the three cosmologies. The Qo'noS and Vulcan results are basically indistinguishable at all redshifts. At $z=0$ the difference between the cosmologies is very small, at $z=2$ the dynamical dark energy model shows an excess of halos at all masses. This is consistent with the results shown in Fig.~\ref{fig:haloc}.} 
\label{fig:halo_mf}
\end{figure}

Next, we show the results for the mass function itself. We focus on the measurements of SO masses, M$_{200c}$. Our halo finder first identifies FOF halos using a linking length of $b=0.168$ and determines the potential minimum for each halo as its center. For FOF halos with more than 500 particles, we measure M$_{200c}$.  In Fig.~\ref{fig:halo_mf} we show the results for the SO halo mass function at three redshifts, $z=0,1,2$. The error bars provided are simply Poisson error bars. The two $\Lambda$CDM cosmologies, with and without neutrinos, are basically identical at all three redshifts and over the full halo mass range. The Ferenginar simulation has more halos in particular at early redshifts. At $z=0$, the halo mass functions of all three simulations are very similar. The finding for the Fereginar simulation, a higher halo mass function at early times and less of a difference at $z=0$ compared to the $\Lambda$CDM simulations, is consistent with the large-scale structure measurements we have shown so far. The slowing-down of structure formation over time is due to a dark energy component that sets in later for the Ferenginar simulation.

\begin{figure}[t]
\centerline{\includegraphics[width=3.5in]{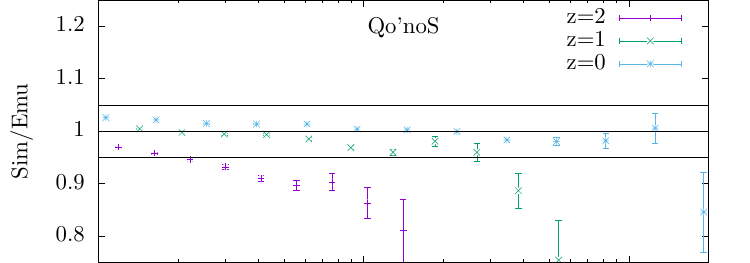}}
\centerline{\includegraphics[width=3.5in]{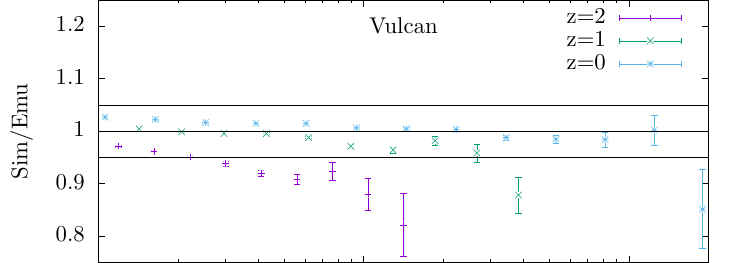}}
\centerline{\includegraphics[width=3.5in]{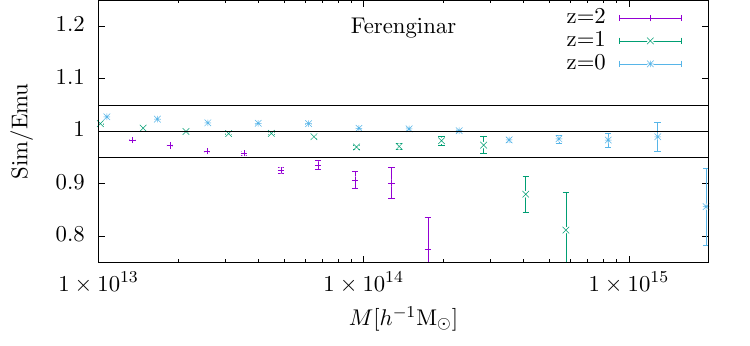}}
\vspace{0.1cm}
\caption{Comparison of the simulation results to the CosmicEmu introduced in \cite{emu_massf}. The solid lines indicate 5\% uncertainty. We only show mass bins with at least 100 halos. The upper panel shows the comparison of the emulator to the Qo'nos simulation, the middle panel the Vulcan results and the last panel shows the Ferenginar results. In all three cases, three redshifts are shown (purple: $z=2$, light blue: $z=1$ and light green: $z=0$).  The best agreement is achieved for $z=0$ at better than 2-3\% for most of the  mass range considered. The agreement degrades with higher redshifts ($\sim$10\% at $z=2$) which is in agreement with the findings in \cite{emu_massf}. We use Poisson error bars.} 
\label{fig:mf_emu}
\end{figure}

Similar to the power spectrum analysis, we show a comparison of our results with a mass function emulator from \cite{emu_massf} in Fig.~\ref{fig:mf_emu}. The emulator provides predictions for SO masses, $M_{200c}$, for group and cluster-sized objects, starting at a mass of $M_{200c}=10^{13}h^{-1}$M$_\odot$. Accordingly, the $x$-axis in Fig.~\ref{fig:mf_emu} is reduced from the full mass function plot shown in Fig.~\ref{fig:halo_mf}. Results for three redshifts are presented for the comparison and for all three models. We restrict the upper mass end to bins that have at least 100 halos. For $z=0$ we find very good agreement between the simulations and the emulator, at the 2-3\% level. At redshift $z=1$, the agreement is still over most of the mass range better than 5\% (indicated by the solid lines in the figure). The accuracy degrades somewhat for $z=2$. These findings are fully consisent with the accuracy reported in \cite{emu_massf}.

\subsection{Concentration-Mass Relation}

The final example measurement we provide is for the concentration-mass (c-M) relation at three redshifts, $z=0,1,2$. The concentration is obtained by fitting a Navarro-Frenk-White (NFW) profile \citep{nfw1,nfw2} to each halo via 
\begin{equation}
\rho(r)= \frac{\delta\rho_{\rm{crit}}}{(r/r_s)(1+r/r_s)^2}.
\label{eq:nfw}
\end{equation}
Here $\delta$ is a characteristic dimensionless density, and $r_s$ is
the scale radius of the NFW profile. The concentration of a halo is then simply given by $c_{\Delta}=r_{\Delta}/r_s$, where $\Delta$ is the overdensity measured with respect to the critical density $\rho_c$, and $r_{\Delta}$ is the radius at which the enclosed mass is equal to the volume of the sphere times the density. As for the mass function, we choose $\Delta=200$, leading to
$c_{200c}= R_{200c}/r_s$. Details about our fitting approach can be found in \cite{bhattacharya13} and \cite{child}. The error bars shown combine the error contributions from the individual concentration measurements and the Poisson error due to the finite number of halo in each bin. For a detailed discussion, the reader is referred to \cite{bhattacharya13}.

\begin{figure}[t]
\centerline{\includegraphics[width=3.5in]{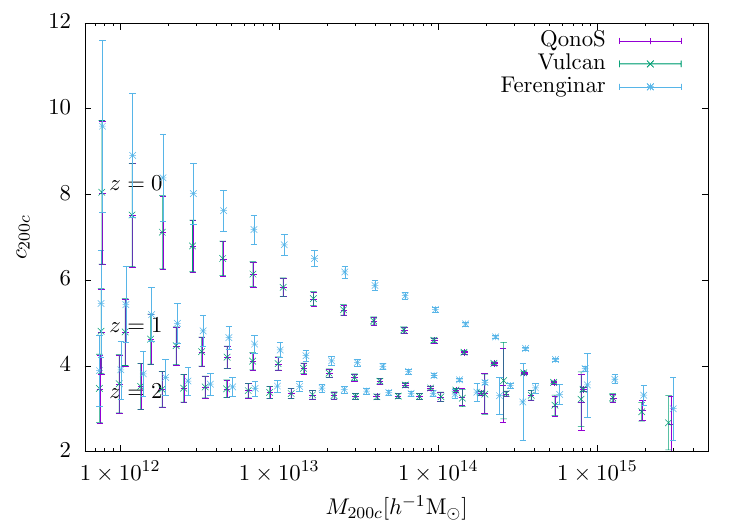}}
\caption{Concentration-mass relation at three different redshifts, $z=0,1,2$ for the three different cosmologies. The difference between the Qo'noS and Vulcan simulations is minimal. At high redshift ($z=2$, left panel), the differences between the three models are small. The Ferenginar simulation shows a more interesting behavior: While at early redshifts, the concentration-mass relation is almost the same as for the $\Lambda$CDM cases, at lower redshifts the concentrations are considerably higher.} 
\label{fig:cM_all}
\end{figure}

The results are shown in Fig.~\ref{fig:cM_all} for the three simulations and three redshifts. The concentration-mass relation for the Qo'noS and Vulcan simulations are basically indistinguishable at all three redshifts. This is to be expected given our assumption that the neutrino mass considered here is small enough to allow us to neglect their nonlinear impact on structure formation (for a more extensive discussion about the effect of neutrinos on group and cluster sized halos see \cite{2019arXiv190110690B}).  

The concentration-mass relation for the Ferenginar simulation shows a different behavior than the $\Lambda$CDM simulations. While at high redshift, $z=2$, all three simulations have a very similar c-M relation, at late times, the dynamical dark energy model exhibits larger concentrations, a $\sim 20\%$ increase, than the other two simulations. The mass function measurements (Fig.~\ref{fig:halo_mf}) show that the halos in the Ferenginar simulation formed earlier. This early formation then translates into tighter, more concentrated, halos at later times. This result is also consistent with the matter power spectrum measurements (Fig.~\ref{fig:pk_nl_z0}) which has more power on small scales in the Ferenginar simulation compared to the other two simulations.

\section{Data Release}
\label{sec:release}

Several HACC simulation outputs have been released via the HACC Simulation Data Portal\footnote{\url{https://cosmology.alcf.anl.gov/}}, described in~\cite{heitmann19b}. The HACC Simulation Data Portal is hosted at the Argonne Leadership Computing Facility (ALCF) and data transfer from the portal is incorporated via Globus\footnote{https://www.globus.org/}. As part of this paper, we release a selected set of outputs from the three simulations\footnote{https://cosmology.alcf.anl.gov/transfer/newworlds}. The data selection is the similar as for previous large simulation releases. We focus on FOF halo properties and downsampled particle files:
\begin{equation}
z=\{0.0, 0.05, 0.21, 0.50, 0.54, 0.78, 0.86, 1.43, 1.49\}.
\end{equation}
For each of the redshifts listed, we provide access to 1\% of the particle data. For each particle, a particle ID, comoving positions $(x, y, z)_{\rm COM}$ in $h^{-1}$Mpc, and 
comoving peculiar velocities $(vx, vy, vz)_{\rm COM}$ in km/s are given. The particles were randomly selected for the first HACC output, but for all following outputs, the same particles are chosen. In addition, halo information is made available. The halos are identified with an FOF halo finder using a linking length of $b=0.168$. For each halo, its center position and velocity are provided. For the center definition, both the gravitational potential minimum and the center of mass measurement are available. For the FOF halo definition, these two quantities can be significantly different, in particular for large, unrelaxed halos.
The halo mass is given in $h^{-1}$M$_\odot$. 

\section{Summary and Outlook}
\label{summary}
In this paper we introduced the ``New Worlds Simulations'', three new gravity-only, extreme-scale simulations. Each simulation covers a volume of (3$h^{-1}$Gpc)$^3$ and evolves 12,288$^3$ particles, making them some of the largest simulations of structure formation currently available. A good value for the mass resolution, $m_p\sim 10^9h^{-1}$M$_\odot$, in combination with the large volume make these simulations particularly useful for investigations related to ongoing and upcoming cosmological surveys. The three simulations have the same initial phases and span three cosmological models, a $\Lambda$CDM cosmology informed by Planck constraints, a $\Lambda$CDM cosmology including massive neutrinos, and a dynamical dark energy model. The dark energy equation of state for the dynamical dark energy model is expressed via the simple $(w_0,w_a)$ parameterization, commonly used by ongoing surveys.

The simulations were carried out with HACC on the Summit supercomputer at OLCF. HACC provides a set of analysis capabilities that are carried out during simulation runtime. In this paper, we showed selected results from this in situ analysis for both particle and halo statistics. The results are consistent with previous findings as demonstrated via comparisons to publicly available emulators. Several interesting differences between the results from the three simulations were discussed in the light of the different physics included in the simulations. Most notably, the varying dark energy equation of state led to stronger clustering at early times, translating for example into higher concentration values and halo counts at higher redshifts. All simulations were normalized to the same value of $\sigma_8$ at $z=0$ and results at the final time are very similar. The addition of a massive neutrino component did not have a large effect on the different statistics we measured due to the small neutrino mass chosen. Overall, the small differences between the measurements we presented in this paper hint at the difficult tasks laying ahead of us to find new fundamental physics, clearly disentangled from systematic effects. The availability of detailed data at different epochs in the evolution of the Universe will be crucial to solidly confirm or reject physics beyond the Standard Model of Cosmology. We publicly release some of our simulation results to enable the creation of synthetic sky catalogs to help the community to further explore the subtle differences between the cosmologies considered in this paper.

\begin{acknowledgments}We thank Kelly Moran and Sebastian Bocquet for discussions about the emulator comparisons. Argonne National Laboratory's work was supported under the U.S. Department of Energy contract DE-AC02-06CH11357. This research used resources of the Oak Ridge Leadership Computing Facility at the Oak Ridge National Laboratory, which is supported by the Office of Science of the U.S. Department of Energy under Contract No. DE-AC05-00OR22725. This research used resources of the Argonne Leadership Computing Facility, a U.S. Department of Energy (DOE) Office of Science user facility at Argonne National Laboratory and is based on research supported by the U.S. DOE Office of Science-Advanced Scientific Computing Research Program, under Contract No. DE-AC02-06CH11357.
\end{acknowledgments}

\end{document}